\documentclass[%
 prl, twocolumn,
superscriptaddress,
 amsmath,amssymb,
 aps,
nofootinbib
]{revtex4-2}

\usepackage{graphicx}
\usepackage{dcolumn}

\usepackage{bm}

\usepackage{xcolor}
\usepackage{xr}
\usepackage{comment}
\usepackage{hyperref}
\usepackage{cleveref}
\usepackage{enumitem}
\usepackage{accents}
\hypersetup{colorlinks,linkcolor={blue!90!black},citecolor={blue!90!black},urlcolor={blue!90!black}}

\DeclareMathAlphabet\mathbfcal{OMS}{cmsy}{b}{n}

\usepackage{amsmath}
\usepackage{mathtools}

\newcommand{\boldnab}{\boldsymbol{\nabla}}
\begin{document}

\title{Anti-Diffusion in an Algae-Bacteria Microcosm:\\ Photosynthesis, 
Chemotaxis, and Expulsion}

\author{Praneet Prakash }
\email[]{pp467@cam.ac.uk}
\thanks{Joint first author}
\affiliation{Department of Applied Mathematics and Theoretical 
Physics, Centre for Mathematical Sciences,\\ University of Cambridge, Wilberforce Road, Cambridge CB3 0WA, 
United Kingdom}
\author{Yasa Baig}
\email[]{ymb27@cam.ac.uk}
\thanks{Joint first author}
\affiliation{Department of Applied Mathematics and Theoretical 
Physics, Centre for Mathematical Sciences,\\ University of Cambridge, Wilberforce Road, Cambridge CB3 0WA, 
United Kingdom}%

\author{Francois J. Peaudecerf}
\email[]{francois.peaudecerf@univ-rennes1.fr}
\affiliation{Institute de Physique de Rennes, Universite Rennes, UMR 6251, F-35000 Rennes, France}%
\author{Raymond E. Goldstein}
\email[]{R.E.Goldstein@damtp.cam.ac.uk}
\affiliation{Department of Applied Mathematics and Theoretical 
Physics, Centre for Mathematical Sciences,\\ University of Cambridge, Wilberforce Road, Cambridge CB3 0WA, 
United Kingdom}%

\date{\today}

\begin{abstract}
In Nature there are significant relationships known between microorganisms from two 
kingdoms of life, as in the supply of vitamin B$_{12}$ by bacteria to algae.  
Such interactions motivate general investigations into the spatio-temporal dynamics of metabolite exchanges.  Here we study by experiment and theory
a model system: a coculture of the bacterium {\it B. subtilis}, an obligate aerobe that is chemotactic to oxygen,
and a nonmotile mutant of the alga {\it C. reinhardtii}, which photosynthetically produces oxygen when illuminated.
Strikingly, when a shaft of light illuminates a thin, initially uniform suspension of the two,
the chemotactic influx of bacteria to the photosynthetically active region
leads to expulsion of the algae from that 
area.  This 
effect arises from algal transport due to spatially-varying collective behavior of bacteria, and is mathematically
related to the
``turbulent diamagnetism" associated with magnetic flux expulsion in stars. 
\end{abstract}
\maketitle

In the early 1880s the biologist Theodor Engelmann performed experiments
that were perhaps the first to use bacteria as sensors \cite{Engelmann1882,Drews,GoldsteinPT}.  
Several
years prior he made the first observation of bacterial chemotaxis
toward oxygen, by showing that putrefactive bacteria would migrate toward the 
chloroplasts of the filamentous alga {\it Spirogyra}.  He then 
determined the ``action spectrum" of 
photosynthesis\textemdash the wavelength-dependent rate of
photosynthetic activity\textemdash by passing sunlight through
a prism and projecting the spectrum onto a 
filamentous green algae held in a chamber that 
contained those self-same bacteria, which  
gathered around the algae in
proportion to the local oxygen concentration, providing
a direct readout of the oxygen production rate.

Although Engelmann's system was engineered for a particular purpose,
and at first glance involves a {\it one-way} exchange of oxygen
for the benefit of bacteria,
there are many examples of mutualistic
exchanges between microorganisms from two distinct kingdoms 
of life.  One of the most significant is that involving vitamin
B$_{12}$.  In a landmark study \cite{B12}, it was shown
that a significant fraction of green algae that require 
this vitamin for their metabolism do not produce it, and
as the ambient concentration of B$_{12}$ in the aqueous
environment is so low, they instead acquire it from
a mutualistic relationship with bacteria, which 
benefit from a source of carbon. 

The study of B$_{12}$ transfer raises fascinating questions in 
biological physics related to the interplay of metabolite
production, chemotaxis, and growth \cite{Peaudecerf},
including the issue of how organisms find each other
and stay together in the turbulent environment of the 
ocean, and how advection by 
fluid flows arising from microorganism motility affects
such mutualisms.
As it is difficult to control the production of vitamin B$_{12}$, we sought to construct a system in which 
the production of a chemical species needed by one member of 
an interacting pair of organisms could be controlled by
the experimentalist.  Taking motivation both from 
Engelmann's experiments and
the B$_{12}$ system, we introduce here a 
coculture \cite{FJPthesis} in which
an obligate aerobic bacterium (one that requires oxygen)
that is chemotactic toward oxygen, coexists with a 
green alga whose photosynthetic activity can be turned 
on and off simply by controlling the 
external illumination.

We use the bacterium \textit{Bacillus 
subtilis}, whose aerotaxis has been central in the
study of bioconvection \cite{PedleyKessler} and in
the discovery of ``bacterial turbulence" 
\cite{Dombrowski}, the dynamical state of a concentrated suspension 
with transient, recurring vortices and jets of collective swimming
on scales large compared to the individual bacteria. 
The alga species is the well-studied unicellular 
\textit{Chlamydomonas reinhardtii}, a model organism 
for biological fluid dynamics \cite{ARFM} with 
readily available motility mutants useful in 
probing the role of swimming in metabolite transfer.
Together these define what we term the 
{\it algae-bacteria-chemoattractant system} (ABC).
While coupled population dynamics
problems have been studied from the familiar reaction-diffusion
point of view in bacterial range expansion 
\cite{mutualrxndiff1}, 
marine \cite{mutualrxndiff2} and more general ecological 
contexts \cite{mutualrxndiff3}, we show 
that there are physical effects that go beyond that standard level of treatment.
Chief among them is the way in which collective 
motion can act
as a ``thermal bath" in enhancing the diffusivity of 
suspended particles \cite{WuLibchaber,Leptos},
which in turn raises fundamental issues concerning 
generalizations of Fick's law \cite{Schnitzer,Lau}.

\begin{figure*}[t]
\centering
\includegraphics[trim={0 0cm 0 0cm}, clip, width=1.5\columnwidth]{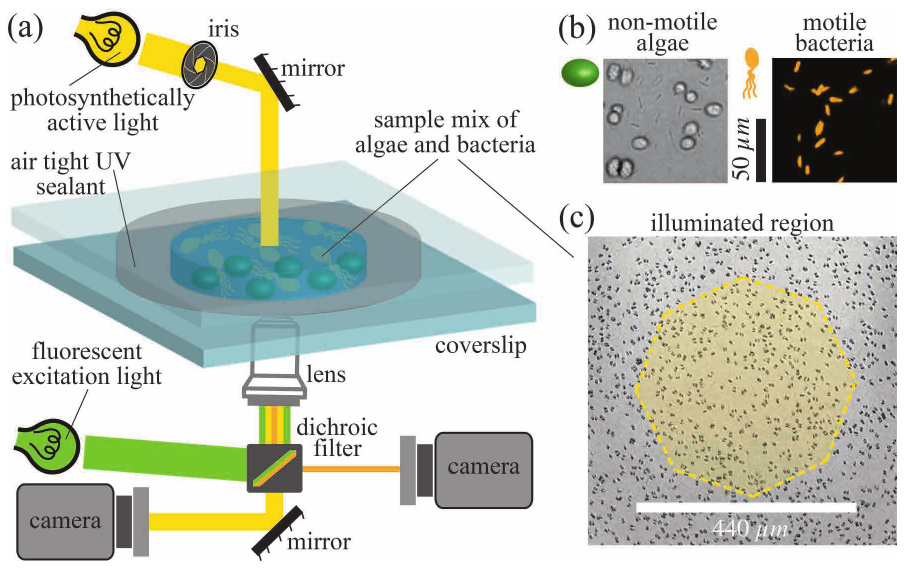}
	\caption{Experimental setup. (a) Schematic of system 
 to illuminate a coculture of a fluorescent strain of the bacterium \textit{B. subtilis} and a 
 flagella-less mutant of the green alga
 \textit{C. reinhardtii}, initially distributed uniformly within the sample chamber.
After a shaft of visible light illuminates the central region the algae produce oxygen to 
which the bacteria are attracted. (b) Algae and bacteria as observed through the brightfield and 
fluorescent channels, respectively. (c) Yellow shading indicates the illuminated region, with 
dark algae visible in the background.}
\label{fig1}
\end{figure*}

\begin{figure*}[t]
\centering
\includegraphics[trim={0 0cm 0 0cm}, clip, width=1.7\columnwidth]{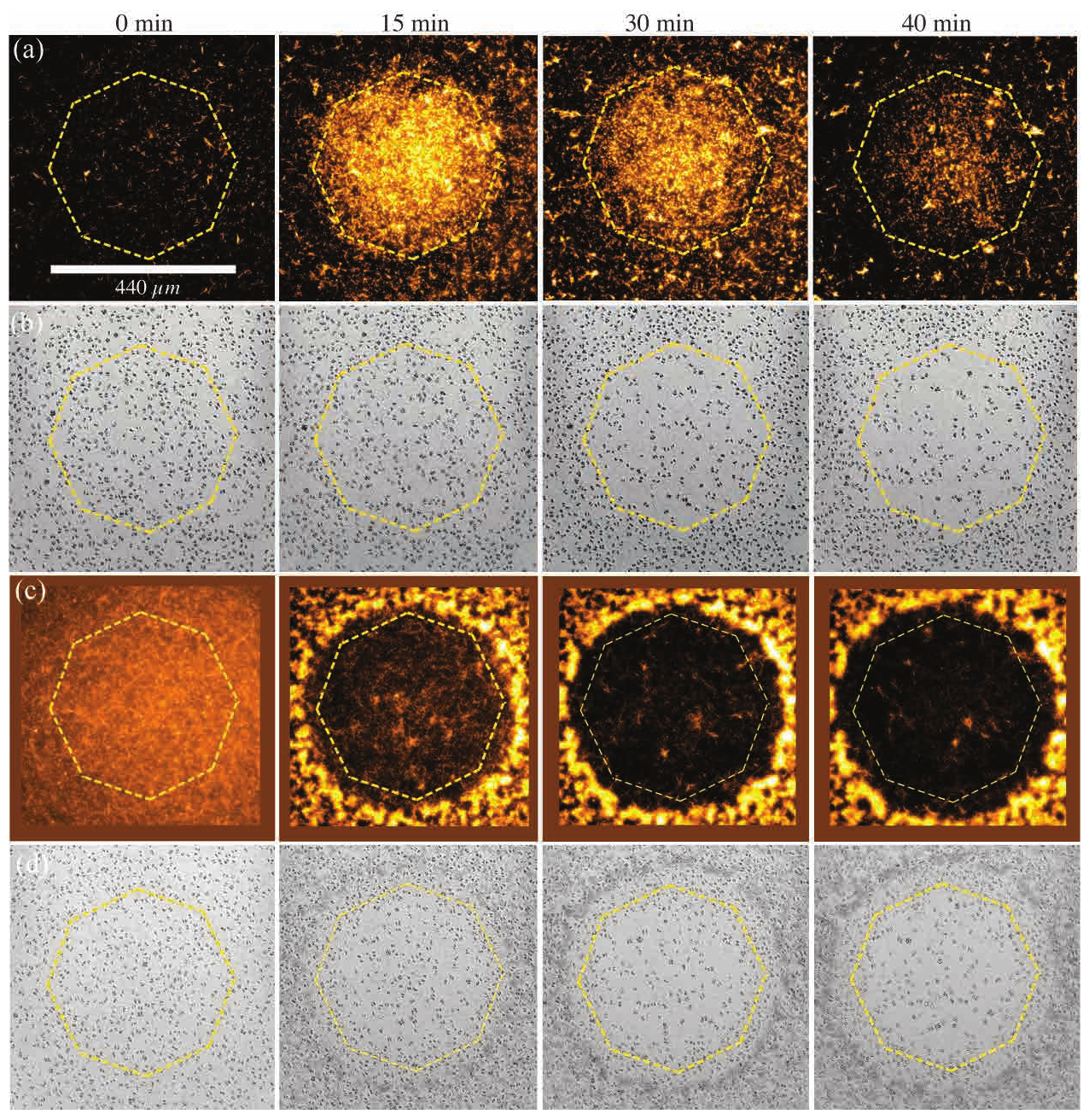}
	\caption{Bacterial influx and algal expulsion. (a) Type-I dynamics.
 Spatio-temporal evolution of bacterial concentration after 
illumination is initiated, where octagonal ring indicates boundary
of illuminated region. At the low initial bacteria concentration ($\sim \! 1\times10^8$ cm$^{-3}$), 
bacteria first move into the illuminated region and then retreat. (b) Algal concentration as a function of
time, demonstrating expulsion from illuminated region. (c) Type II dynamics. 
At a higher concentration ($\sim\! 5\times10^8$ cm$^{-3}$), many bacteria become non-motile and are expelled into the dark region, 
forming a concentrated circular accumulation 
that acts as a natural boundary. (d) Algae are also expelled, accumulating 
inside the confines of the bacterial accumulation.}
\label{fig2}
\end{figure*}

A disarmingly simple experiment to understand the
dynamics of the ABC system involves a thin, 
quasi-two-dimensional suspension of non-motile algae and fluorescently labelled 
bacteria at initially uniform concentrations.  
As depicted 
in Fig. \ref{fig1}, a shaft of photosynthetically active light is cast on the suspension,
triggering oxygen production by the illuminated algae.  
As shown in Figs. \ref{fig2}(a,b), this leads to chemotaxis of bacteria into the 
illuminated region, producing a high 
concentration of bacteria.
Remarkably,
we find that algae are then expelled 
from the illuminated region. Quantitative measurements of the 
local bacterial dynamics in the the system 
show that this expulsion is associated with a 
{\it gradient of collective bacterial behavior} from its peak at
the center, leading to a
outward algal transport.  On longer time scales,
after the algal expulsion, the bacterial concentration
returns to uniformity as the bacteria diffusive 
outward in the absence of chemotactic stimulus.  We name this 
{\it Type I} dynamics. 
Figures \ref{fig2}(c,d) show that at sufficiently high initial bacterial concentrations a new, {\it Type II} behavior is observed; 
expulsion of alga and
consumption of oxygen are sufficiently rapid
that many bacteria in the illuminated region become hypoxic, transition to an 
immotile state, and are
then also expelled into the dark. 

We develop here a system of coupled PDEs that
describes the ABC system 
that provides a quantitative account of the
experimental observations.  These PDEs incorporate
diffusion, chemotaxis, oxygen production and
consumption and algal transport by collective
effects.  The expulsion of algae is 
similar to well-known 
processes
in magnetohydrodynamics termed ``flux expulsion".  First 
discovered for magnetic fields in 
a prescribed constant vortical 
flow field \cite{Weiss}, in which field 
lines are expelled from the vortex, it was later recognized that this expulsion can arise from 
gradients in the intensity of turbulence whose
random advection of the magnetic vector potential  
leads to a gradient in its effective diffusivity 
and thence to ``turbulent diamagnetism" 
\cite{Tao}.

{\it Methods.\textemdash}  
The coculture used strain 168 of {\it B. subtilis} which was 
genetically engineered to express yellow fluorescent protein 
m-Venus with excitation at $515\,$nm and emission at $528\,$nm \cite{Steiner}. 
A single bacterial colony was picked from an agar plate 
and grown overnight in Terrific Broth (TB) on an 
orbital shaker at $240\,$rpm and $30\,^{\circ}$C. The bacteria intended for experiments 
were grown from overnight culture until 
exponential growth phase in Tris-min medium spiked 
with TB medium (Tris-min $+$ $0.1\,$ \% w/v glycerol $+$ 
$5\,$ \% w/v TB). The non-motile 
{\it C. reinhardtii} strain CC477 ({\it bld1-1})
was sourced from the Chlamydomonas Resource Center 
\cite{CRC} and grown in Tris-min medium, 
on an orbital shaker at $240\,$rpm and $20\,^{\circ}$C. The diurnal cycle 
was $12\,$h cool white light ($\sim 15 \, \mu$mol  
photons/m$^2$s  PAR), and $12\,$h in the dark. 

Prior to experiments, bacteria and alga  
from their respective exponential growth phases were mixed in a 
modified Tris-min medium, with glycerol as a 
carbon source and bovine serum albumin to 
prevent cell adhesion (Tris-min $+$
$0.1\,$\% w/v glycerol $+$ $0.01\,$\% v/v BSA). 
The desired number density 
of cells was achieved by centrifugation at $4000\,$g. All cell concentrations 
and 
sizes were measured using a Beckman Coulter Counter 
(Multisizer 4e). The concentration of algae was fixed 
at $5\times10^6\,$ cm$^{-3}$ while varying the bacteria 
concentration in the range $(0.5-5)\times10^8$ cm$^{-3}$. 
The mixture of cells for each experiment 
was transferred to a glass cover slip chamber with a depth 
of $300\,\mu$m, separated by double-sided tape, and 
sealed airtight using UV glue. The chamber surfaces 
were passivated with PEG 
($M_w = 5000\,$ g/mol). 

Experiments were performed on a Nikon 
TE2000-U
inverted microscope. The spatio-temporal variation in 
bacterial concentration 
was monitored by epifluorescence illumination with a 
$\times 20$ objective using a highly sensitive, 
back-illuminated camera (Teledyne Prime $\Sigma$ 95B). 
Movies of algae cells were recorded through the 
brightfield channel using a Phantom V311 high-speed camera 
(Vision Research) at $\times20$ magnification. The halogen lamp served as 
both a brightfield 
and photosynthetic light source. The size of a light shaft used
to trigger photosynthesis was controlled by the field 
iris in the microscope condenser arm,
producing an octogonal boundary in the focal plane with 
mean radius of $R=220\,\mu$m.

\begin{figure*}[t]
\centering
\includegraphics[trim={0 0cm 0 0cm}, clip, width=1.8\columnwidth]{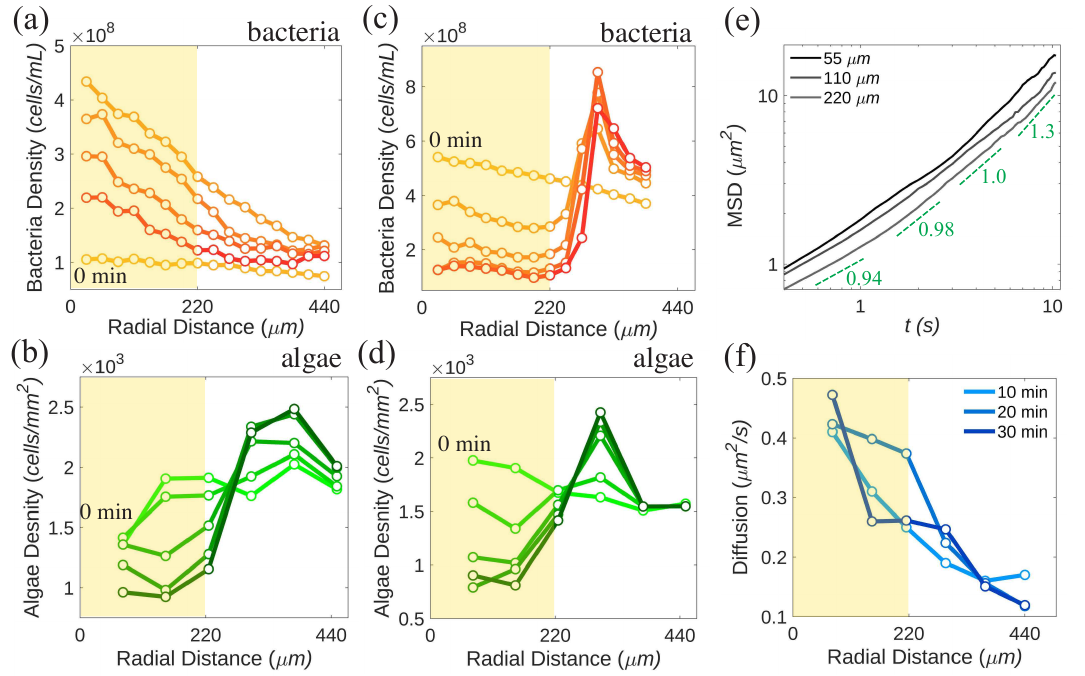}
e\caption{Bacterial accumulation and algal 
expulsion. Yellow shading in the plots up to $220\, \mu$m indicates illuminated region. 
(a) Bacterial and (b) algal concentrations for a coculture with initial 
concentrations $b=1\times 10^8$ cm$^{-3}$ and $a=5\times 10^8$ cm$^{-3}$, 
showing Type I dynamics. 
Shading of symbols and lines increase with time, with a ten minute interval between data sets.
(c,d) As in (a,b) but with initial bacterial concentration $b=5\times10^8$ cm$^{-3}$, exhibiting Type II dynamics. (e) MSD of algae versus time for the case (a,b) at different radii in the illuminated region.  Dashed lines indicate apparent slope at different 
times.  (f) Algal diffusivity determined in linear regime of MSD versus radial distance at various times after start of illumination.}
\label{fig3}
\end{figure*}

{\it Experimental Results.\textemdash}
Figures \ref{fig2} and \ref{fig3} summarize the
main experimental observations
associated with a homogeneous initial
condition in darkness that is then
illuminated with a shaft of light.
For an initial bacterial concentration of $b=1\times 10^8\,$cm$^{-3}$ 
giving Type I dynamics, 
we observe over the course of the first $\sim 10\,$min after the start 
of illumination that the concentration of bacteria at the center of the illuminated region 
increases dramatically, as
visualized in Fig.~\ref{fig1}(a) and quantified in Fig. \ref{fig2}(b), reaching a
peak enhancement of a factor of $\sim 5$, with
a roughly linear decrease out to the edge.
That peak then relaxes away until the bacterial 
concentration is again nearly uniform after $\sim 35\,$min.
At the peak of accumulation, after $\sim 15\,$min,
there is a clear bacterial depletion zone just outside the
illuminated region, whose width is estimated to be 
$\sim\!150\,\mu$m.
During this period, as shown in Fig. \ref{fig2}(b), the algal
concentration becomes strongly depleted in the illuminated
region, leading to a ring of accumulation at the boundary.

We measured the mean squared
displacement (MSD) of algae versus time at a range of radii $r$ 
from the light shaft center.  The MSD in Fig. \ref{fig3}(e) exhibits systematic 
upward curvature, with a local exponent of unity at time $t_1\sim 6-7\,$s, 
but faster
behavior for $t > t_1$.  Such superdiffusion for tracers is a well-known 
consequence of active turbulence in concentrated bacterial 
suspensions \cite{Dombrowski,WuLibchaber}. 
A heuristic illustration of the strong gradient in collective behavior in the 
illuminated region is obtained by determining an effective algal diffusion 
constant $\tilde{D}_a$   
from the slope of the MSD curve at $t_1$ (Fig. \ref{fig2}(f)), which
is a strongly decreasing
function of distance from the shaft center, that mirrors the decreasing 
bacterial
concentration.  
The largest of these diffusivities is $\sim 25$ times the 
purely thermal value 
$D_{\rm th}=k_BT/6\pi\mu a\simeq 0.02\, \mu$m$^2$/s, where 
$\mu$ is the medium viscosity and $a=10\,\mu$m is 
twice the algal radius since most algae exist as 
pairs (``palmelloids").   Yet, by itself, 
an effective algal diffusivity $\tilde{D}_a\sim 0.5\,\mu$m$^2$/s implies
a diffusive time $R^2/\tilde{D}_a\sim 700\,$min for algae at the center 
to escape purely by random motion, far 
longer than the observed time of $30\,$min; 
the superdiffusive behavior
seen for $t>t_1$ in Fig. \ref{fig3}(e) 
signals a qualitatively different transport
process. 

At the higher initial bacterial concentration $b=5\times 10^8\,$cm$^{-3}$ 
we observe the distinct Type II 
behavior: expulsion happens much more rapidly, the
bacterial concentration profile becomes nonmonotic in radius, with a 
strong peak just outside the 
illuminated region, and the peak of the concentration of expelled 
algae is narrower.  Close microscopic inspection of the region of high
bacterial concentration in the dark region shows that most of the bacteria
there are immotile and the expelled algae reside just inside the
bacterial accumulation ring.  We deduce that many bacteria in the 
illuminated region have become hypoxic and are expelled from that area
through much the same process that expelled the algae.

\begin{figure*}[t]
\centering
\includegraphics[trim={0 0cm 0 0cm}, clip, width=1.8\columnwidth, scale = 0.6]{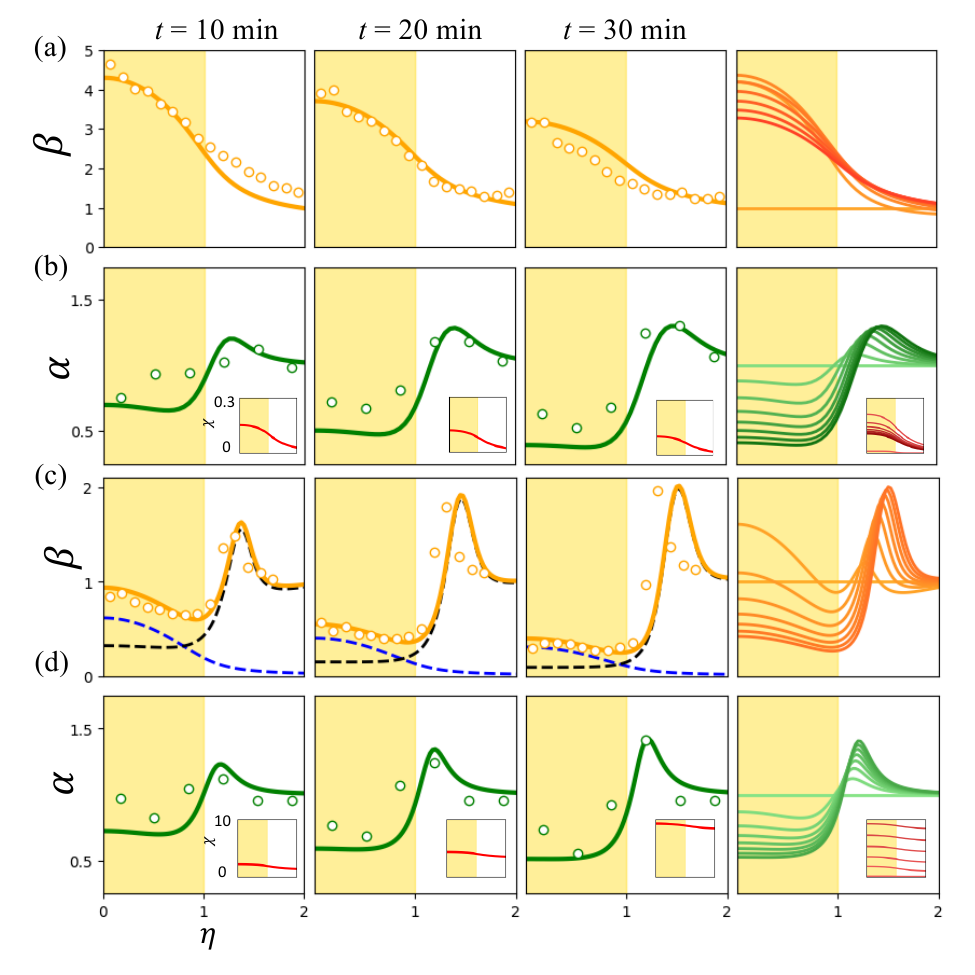}
\caption{Theoretical predictions. (a,b) Type I dynamics. (a) Non-dimensional bacterial dynamics 
from experiment (circles) compared to theoretical fit (solid lines) for first thirty minutes of
experiment. Rightmost column in each row shows the theoretical time evolution over 30 minutes, sampled every $\sim$ 3 minute. Shading intensity increases with time. (b) As in (a) but for 
algae, showing expulsion. Insets show theoretical oxygen profiles $\chi$. 
(c,d) Type II dynamics. (c) Experimental bacterial dynamics (circles) compared to 
predictions of the ABCD model. Dashed lines indicate immotile (black) and active (blue) 
bacteria, whose sum is orange solid line. Rightmost plot
shows time evolution of the total bacterial concentration. (d) Algal
expulsion for high density experiment. For (a),(b) we took $\kappa^2 = 1, 
\epsilon = 1, d_b = 0.1, d_a = 0.0001, \gamma = 1.2, \zeta = 0.002$. For (c), (d), we use $ \gamma = 0.35, \zeta = 0.005, \kappa^2 = 5, d_{\delta} = 10^{-4}, \rho = 1.0 , \chi^* = 0.1, \zeta_d = 0.04, \delta^* = 0.01$ with all other parameter values held over.}
\label{fig4}
\end{figure*}

{\it The ABC model.\textemdash}
We now turn to a mathematical model for the behavior described above, 
focusing first on Type I expulsion.
In this case, a minimal description of the system entails 
the concentration 
fields $a({\bf r},t),b({\bf r},t)$ and $c({\bf r},t)$ for 
algae, bacteria, and oxygen, respectively. 
If $D_c$ is the oxygen diffusion constant (assumed independent 
of all other variables), $k_+$ is the rate of oxygen
production per algal cell when illuminated with a
spatially varying light intensity $I(r)$ and the oxygen consumption rate per 
bacterium has a Michaelis-Menten form, then
\begin{equation}
    c_t=D_c\nabla^2 c + k_+ a I(r)-k_- b\frac{c}{K_c+c},
    \label{odynamics}
\end{equation}
where $K_c$ is the Michaelis constant.

Consider first the situation of uniform illumination 
($I=1$) and uniform concentrations $a_0$ and $b_0$ of 
algae and bacteria.  A steady state $c^*=K_c k/(1-k)$, 
with $k=k_+a_0/k_-b_0$, can be reached in which 
consumption balances production, provided $k<1$.
We assume this inequality is always satisfied 
and typically invoke the {\it weak  production limit}
$k\ll 1$, in which the 
oxygen consumption can be approximated as 
$k_- bc/K_c$.

We explain algal expulsion from the
illuminated region through three 
intermediate calculations; oxygen production in a 
uniform suspension; bacterial chemotaxis in the
presence of oxygen production; algal dynamics due
to an inhomogeneous bacterial concentration.
Suppose that the light is constrained to a shaft of
radius $R$.  Scaling time, space, and concentrations 
via $T=tD_c/R^2$, $\eta=r/R$,
$\chi=c/c^*$, $\alpha=a/a_0$, and $\beta=b/b_0$, letting 
$\epsilon=k/(1-k)$, and introducing
the {\it screening length} 
\begin{equation}
    \lambda=\left(D_c\tau_c\right)^{1/2},
\end{equation}
where $\tau_c=K_c/k_-b_0$ is the characteristic consumption time of oxygen, 
the dynamics \eqref{odynamics} takes the
form 
\begin{equation}
    \chi_T=\nabla^2\chi+(1-k)\kappa^2\alpha\Theta(1-\eta)-\kappa^2\beta\frac{\chi}{1+\epsilon\,\chi},
    \label{scaled_odynamics}
\end{equation}
where now $\nabla^2=\nabla^2_\eta$,  and 
$\Theta$ is the Heaviside function.
Here, $\kappa^2=R^2/\lambda^2$ is also the ratio $\tau_D/\tau_c$
of the time $\tau_D=R^2/D_c\sim 25\,$s for oxygen to 
equilibrate diffusively across the illuminated region to
the consumption time. 

If we clamp concentrations $\alpha$ and 
$\beta$ at unity, take $k\ll 1$, and enforce continuity in $\chi$ and $\chi_\eta$ at 
$\eta=1$, the steady state of \eqref{scaled_odynamics} in an
unbounded domain is 
\begin{equation}
    \chi(\eta)=\begin{cases}
        1-\kappa K_1(\kappa)I_0(\kappa\eta) & \eta\le 1,\\
        \kappa I_1(\kappa)K_0(\kappa \eta) & \eta\ge 1,
    \end{cases}
    \label{chi_ss}
\end{equation}
in terms of modified Bessel functions $K_\nu$ and 
$I_\nu$.  From the inner solution we see that the oxygen concentration at the
centre of the illuminated region asymptotes to unity ($c^*$ in unrescaled units) 
for large domain size $\kappa$, but is attenuated strongly as $\kappa$ falls below unity.
The outer solution
behaves as $\chi\sim 
\exp(-(r-R)/\lambda)$, 
showing that $\lambda$ serves
as the characteristic
penetration depth of oxygen into the surrounding
bacterial population and sets the depletion zone seen
in Fig. \ref{fig1}(b).  With $D_c\sim 2\times 10^{3}\,\mu$m$^2$/s and $R=220\,\mu$m, and the estimate $\tau_c\sim 10^2\,$s \cite{pnas}, 
we find $\lambda\sim 400\,\mu$m, and thus $\kappa\sim 1-2$.

Relaxing the assumption of uniform bacterial concentration, it is clear that beyond the time 
$\tau_D$ bacteria within a 
distance $\lambda$ of the edge of the illuminated 
region will experience the steepest oxygen gradient
and chemotax most rapidly inwards.  This can be
described by the simplest combination of
diffusion and chemotaxis as in the Keller-Segel model,
\cite{KellerSegel},
$b_t=D_b\nabla^2 b -\boldsymbol{\nabla}\cdot
    \left(g b\boldsymbol{\nabla}c\right)$
where the diffusion constant $D_b$ arises from random 
cellular swimming and the response coefficient $g$ is taken to be
constant.   
Scaling as above, we obtain 
\begin{equation}
    \beta_T = d \nabla^2 \beta - \gamma \boldnab \cdot (\beta \boldnab \chi),
    \label{scaled_bdynamics}
\end{equation} 
where $d = D_b/D_c$, and $\gamma = gc^*/D_c$. 

A steady state can be reached when the chemotactic 
flux $\gamma \beta \boldnab \chi$ balances 
the diffusive flux $d \boldnab \beta$, 
yielding 
\begin{equation}
\beta_{ss}(\eta) = A \exp\left( \dfrac{\gamma}{d} \chi(\eta) \right), 
\label{betass}
\end{equation}
where $A$ is a normalization constant.
While this steady state profile captures the observation that the bacterial concentration 
reaches its maximum at $\eta=0$, where $\chi$ itself
is maximized, the steady state profile 
\eqref{betass} only
develops on time scales sufficient for bacteria 
outside the depletion zone to move inwards and
replenish partially the depletion.  This time will
be at least $(R+\lambda)^2/D_b\gg \tau_D$.  Prior
to this the bacterial concentration is nonmonotonic,
with the depletion zone seen in Fig. \ref{fig2}(a).

The flagella-less algae used in our experiments do not swim;
their movement arises from collective flows driven
by the concentrated bacteria.  At the very least this leads to enhanced 
diffusivity, which, as seen in Fig. \ref{fig3}(f), varies in space, and by itself, 
is captured by an algal flux ${\bf J}_D=-\tilde{D}_a{\boldsymbol \nabla}a$.  
The appearance of ``bacterial turbulence" at higher concentrations implies that the algae
are passive scalars in an {\it inhomogeneous} turbulent flow.  It is well-known that passive particles will be
expelled into separatrices between regions of high vorticity.   When those boundaries change with time, the effect to transport particles from regions of high
turbulence to low, without requiring any gradients in particle concentration, akin to
{\it chemokinesis}, where cells accumulate in regions where they swim slowly.  In a simple approximation, the extent of the
turbulent transport is proportional to the bacterial concentration, giving a 
contribution to the algal flux ${\bf J}=-p a{\boldsymbol \nabla}b$ for
some $p>0$.  With the concentration seen in Fig. \ref{fig3}(a), whose
radial gradient is negative, this contribution leads to an {\it outward} flux
of algae.

Assembling these contributions and rescaling as above, we obtain two equivalent
forms of the dynamics,
\begin{subequations}
\begin{align}
\alpha_T &= \boldnab\cdot\left(d_{a} \boldnab\alpha\right)
+\zeta\boldnab\cdot\left(\alpha\boldnab\beta\right),\label{aeom1}\\
\alpha_T &- \zeta \boldnab \beta  \cdot \boldnab \alpha = 
\boldnab\cdot\left(d_{a} \boldnab\alpha\right)
+ \zeta \alpha\nabla^2\beta,\label{aeom2}
\end{align} 
\end{subequations}
where $d_a=\tilde{D}_a/D_c$ and $\zeta= pb_0/D_c$.  
In the first form \eqref{aeom1}, we see a parallel to the bacterial chemotaxis 
equation \eqref{scaled_bdynamics}; algae exhibit negative 
``bacteria-taxis".  
In symmetrizing the two problems, this ``completion" of the dynamics is loosely
analogous to the introduction of the displacement current in Maxwell's equations 
as a parallel to Ampere's law.
In the second form \eqref{aeom2}, we see that there is an explicit advective contribution 
and a ``reactive" term 
$\alpha\nabla^2\beta$.  This latter contribution plays an important role in the algal expulsion, as
the bacterial concentration within the illuminated region has a large negative second 
derivative, and this term thus forces $\alpha$ down there, whereupon it is 
advected outward.  This is a process of ``anti-diffusion".

Figures \ref{fig4}(a,b) show how the model defined by Eqs. \eqref{scaled_odynamics}, 
\eqref{scaled_bdynamics} and \eqref{aeom2} provide a 
quantitative
fit to the observed Type I dynamics.  In particular, we see the prompt accumulation of
bacteria in the illuminated region followed by expulsion of algae.  In the 
model and in experiment, the
fraction of algae ultimately expelled is $\sim 0.5$, so there is still oxygen 
production within the illuminated region after $\sim 30\,$min, leading to
continued chemotactic attraction of bacteria inward.  This leads to a slower
decay of the bacterial concentration back to its original uniform, low value than
seen in experiment.  This may reflect processes such as bacterial adaptation
to the oxygen and a gradual reduction in oxygen production by algae.
 
{\it The ABCD model.\textemdash}
While the ABC model can account for the essential features of Type I dynamics, it does not
allow for loss of motility of bacteria at low oxygen concentrations in Type II 
dynamics.  This transformation
has been recognized as important in the context of bioconvection \cite{pnas,Hillesdon},
where the influx of oxygen at the air-water interface of a bacterial suspension competes with
consumption within the fluid, leading to a hypoxic region hundreds of microns below the surface.  Hypoxia-induced motility transitions have also been observed in
the penetration of oxygen into suspensions of {\it E. coli} \cite{Douarche}.

To account for this transformation we view 
the ``dormant", non-motile state of the bacteria, with 
concentration $d$, as a separate population distinct
from the motile form, so the extended ``ABCD" model involves 
algae, bacteria, chemoattractant, and dormant bacteria.
The interconversion rate as a function of oxygen concentration 
$c$ is taken as a simple generalization 
of the substrate-dependent growth of the Monod model,
$v_{con}b K_{sat}/(c+K_{sat})$, where $v_{con}$ is the maximum 
conversion rate to the immotile 
form and $K_{sat}$ is the concentration at which half-maximal conversion 
occurs. 

The dynamics of dormant bacteria include 
generation, expulsion, and a very small diffusion constant $D_d$.  
Setting $\delta=d/b_0$, the rescaled dynamics takes the form
\begin{equation}
\delta_T - \zeta_d \boldnab \beta  \cdot \boldnab \delta = 
d_d\nabla^2 \delta + \zeta_d \delta\nabla^2\beta+ \rho\beta \frac{\chi^*}{\chi^*+\chi},
\label{delta_eom}
\end{equation}
where $d_d=D_d/D_c$, $\rho=v_{con}R^2/D_c$ and $\chi^*=K_{sat}/c^*$. 
Accordingly the bacterial dynamics \eqref{scaled_bdynamics} acquires the 
corresponding loss term from conversion, becoming
\begin{equation}
    \beta_T = d \nabla^2 \beta - \gamma \boldnab \cdot (\beta \boldnab \chi)- \rho\beta \frac{\chi^*}{\chi^*+\chi}.
    \label{scaled_bdynamics2}
\end{equation} 
A final modification to the ABC model involves steric
effects on algal diffusion that occur when the concentration
of dormant bacteria is large.  This is modelled by modifying
the rescaled algal diffusivity $d_a$ to 
$d_a(1-\delta/(\delta+\delta^*))$, a form that reduces $d_a$ at large
$\delta$ but retains its positivity.
With these components, Fig. \ref{fig4}(c,d) show that the ABCD
model captures the expulsion of both
algae and immotile bacteria, with
a narrow accumulation of dormant bacteria blocking transport of algae.

\begin{figure}[t]
\centering
\includegraphics[trim={0 0cm 0 0cm}, clip, width=1.\columnwidth, scale = 1]{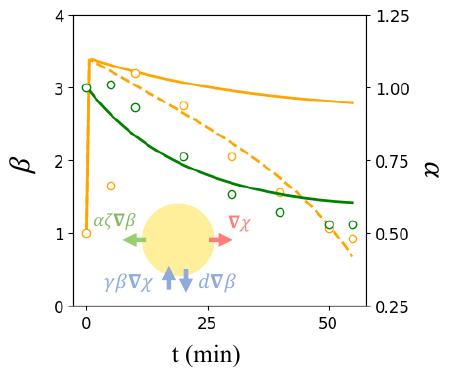}
	\caption{Spatially averaged dynamics.  Average concentration of bacteria (orange) and 
 algae (green) in illuminated region from experiments (circles) and model (lines)
 shown in scattered points. 
 For bacteria, simple 
 chemotaxis is shown by solid orange line, adaptive chemotaxis shown dashed.  Inset: Schematic of 
 fluxes across the illuminated region.}
\label{fig5}
\end{figure}

{\it Spatially-averaged model.\textemdash}
Insight into the basic process of algal expulsion can be obtained
by constructing a spatially-averaged model which takes as dynamical
degrees of freedom the mean concentrations of algae, bacteria, and
oxygen inside the illuminated region, denoted by $\bar{\alpha}$, $\bar{\beta}$, and
$\bar{\chi}$.
As indicated schematically in Fig. \ref{fig5}(a), the specification of this
averaged model requires estimates of the various fluxes across the boundary of
the illuminated region.  
We anticipate that turbulent bacterial dynamics will homogenize oxygen inside
the illuminated region, save
for the area close to the boundary of width $\lambda$ where the concentration 
rapidly decreases.  From an integral form
of \eqref{odynamics} and the divergence theorem we have
$\partial_t(\pi R^2 \bar{c})=-D_c\oint ds {\hat{\bf n}}\cdot \boldnab c$,
where ${\hat{\bf n}}$ is the outward normal to the illuminated region. 
If we estimate ${\hat{\bf n}}\cdot \boldnab c\sim \bar{c}/\lambda$, then
the PDE \eqref{odynamics} becomes the ODE
\begin{equation}
    \frac{d\bar{\chi}}{dT}=-2\kappa \bar{\chi}+(1-k)\kappa^2 \bar{\alpha}
    -\kappa^2 \bar{\beta}\frac{\bar{\chi}}{1+\epsilon \bar{\chi}}.
\end{equation}

Similar estimates hold for the averaged bacteria dynamics, where 
the relevant fluxes are the outward diffusive and inward chemotactic 
contributions.  The diffusive contribution is precisely analogous to
that for oxygen above.  Since on the time scales of the experiment only 
bacteria within a distance $\lambda$ outside the illuminated region
sense the gradient, the inward Keller-Segel flux is also estimated
as above, but with an upper limit on the bacterial concentration due
to steric effects. This leads to the ODE version of \eqref{scaled_bdynamics},
\begin{equation}
    \frac{d\bar{\beta}}{dT}=-2d\kappa \bar{\beta}
    +2\gamma\kappa \left(\beta_{\rm max}-\bar{\beta}\right).
\end{equation}
Finally, the algae have an active flux $-pa\boldnab b$ that leads to 
expulsion, but there is little back-diffusion into the illuminated region.  Yet, 
as algae accumulate at the boundary (Fig. \ref{fig2}), they form a thick ring 
that inhibits continued expulsion of algae. Introducing a 
saturation of the algal flux we obtain the ODE version of \eqref{aeom1},
\begin{equation}
    \frac{d\bar{\alpha}}{dT}=2\zeta\kappa \bar{\beta}\left(\alpha_{\rm max}
    -\bar{\alpha}\right).
\end{equation}

Using parameter values taken from Fig. \ref{fig4} and fitting 
$\beta_{\text{max}} \approx 5$  and $\alpha_{\text{max}} \approx 0.5$, we 
obtain the solid green and orange curves shown in Fig. \ref{fig5} for mean 
algal and bacterial concentrations within the illuminated region.  The 
reduced model captures the slow exponential decay of the algal concentration, but 
while it reproduces the rapid bacterial influx in the first few minutes of the experiments, 
the predicted decay of the bacterial concentration, as remarked earlier for the
full ABC model, is far 
slower than observations. The possibility that this discrepancy arises from adaptation of
the bacteria to elevated oxygen levels can be explored by assuming a simple 
linear decay of the chemotactic coefficient, as $\gamma(T) = \gamma_0 - T/\tau$,
where $\tau$ sets the adaptation rate. For $\tau = 1200$, the bacterial 
dynamics (dashed orange line in Fig. \ref{fig5}) matches closely with data, 
while the algal expulsion (not depicted) is essentially unchanged from the 
non-adaptive case. 

{\it Discussion.\textemdash} 
As our results suggest new phenomenology in active matter systems, it is instructive to return to the connection between algal expulsion and 
flux expulsion in magnetohydrodynamics.  In the original case considered \cite{Weiss}, a magnetic field in a fluid
with velocity ${\bf u}$ obeys the
Maxwell equation,
\begin{equation}
    {\bf B}_t=\boldnab\times\left({\bf u}\times{\bf B}\right)+D_m\nabla^2{\bf B},
\end{equation}
where $D_m$ is the magnetic diffusivity, and both ${\bf B}$ 
and
${\bf u}$ are taken to confined to the $xy$-plane.  The  
magnitude of the magnetic vector potential 
${\bf A}=A\hat{\bf z}$ then obeys the advection-diffusion
equation
\begin{equation}
    A_t+{\bf u}\cdot \boldnab A=D_m\nabla^2 A.
\end{equation}
With ${\bf u}$ a prescribed single-vortex velocity field,
the initial condition $A=B_0x$, corresponding to the space-filling uniform 
magnetic field ${\bf B}=\boldnab\times {\bf A}=-B_0\hat{\bf y}$, homogenizes inside the 
vortex, leaving ${\bf B}=0$ zero there but $A$ essentially undisturbed outside.

In the case of turbulent transport consider later \cite{Tao}, it was shown that on
long time and length scales the effect of the inhomogeneous turbulence is 
captured by an equation of motion for a vector potential component 
$\bar{A}$, averaged on those scales, of the form
\begin{equation}
    \bar{A}_t=\boldnab\cdot\left(D^*\boldnab \bar{A}\right)
\end{equation}
where $D*$ is a turbulent diffusivity tensor. For the particular case 
in which the turbulence varies along one direction, say $y$, then the averaged 
$x$-component of the magnetic field ${\cal F}=\bar{B}_x$ obeys 
\begin{equation}
{\cal F}_t-2(\partial_y D^*)\partial_y{\cal F}=D^*\nabla^2{\cal F}+
{\cal F}\partial_{yy}D^*.
\label{TPW}
\end{equation}
This form has the same structure as our Eq. 
\eqref{aeom1}, with the bacterial concentration $\beta$ playing the role
of the diffusivity tensor component.  We conclude that the
main difference between the two problems is the  
expulsion of a scalar (concentration) field in the 
present context compared to the expulsion of a vector
quantity in MHD.

The results described here 
highlight the rich dynamical behavior that occurs in mixed
active matter systems involving microorganisms
from two Kingdoms of Life.  While the dynamics of symbiosis is generally studied
on time scales relevant to population dynamics or evolution 
\cite{Piskovsky2023, Baig2023, David2013}, our findings indicate
that short-term dynamics on the scale of minutes and hours can have a considerable
impact on the spatial-temporal aspects of association, where chemotaxis and phenotypic
switching dominate.  
Although we considered the simple case of light intensity
that is piecewise constant, inhomogeneous activity arises.
The treatment of such nonuniform active matter has received attention only recently,
in the context of ``invasion" \cite{Miles} and active gels \cite{Assante},
but is central to any discussion of realistic ecologies.
In this sense, generalizing the present setup to allow
algal motility and phototaxis in the presence of an inhomogeneous light field 
may reveal even more striking dynamics when the oxygen sources are 
themselves motile.

\begin{acknowledgments}
We are grateful to Jim Haseloff for providing the fluorescent strain of {\it B. subtilis} 
and to Kyriacos Leptos for numerous discussions. This work was supported in part
by the Gordon and Betty Moore Foundation, Grant No. 7523 (PP \& REG) and the U.K. Marshall Aid Commemoration Commission (YB). 

\end{acknowledgments}



\begin{thebibliography}{99}

\bibitem{Engelmann1882} T.W. Engelmann, Ueber Sauerstoffausscheidung von Pflanzenzellen im Microspectrum,
\href{https://doi.org/10.1007/BF01802976}{Archiv f{\"ur} die gesamte Physiologie des Menschen und der Tiere {\bf 27}, 485-489 (1882)}.

\bibitem{Drews} G. Drews, 
Contributions of Theodor Wilhelm Engelman on phototaxis, 
chemotaxis, and photosynthesis, 
\href{https://doi.org/10.1007/s11120-004-6313-8}{Photosynth. Res. {\bf 83}, 25--34 (2005)}.

\bibitem{GoldsteinPT} R.E. Goldstein, 
Coffee stains, cell receptors, and time crystals: Lessons from
the old literature, 
\href{https://doi.org/10.1063/PT.3.4019}{Phys. Today {\bf 71},  32--38 (2018)}.

\bibitem{B12} M.T. Croft, A.D. Lawrence, E. Raux-Deery, 
M.J. Warren, and A.G. Smith,
Algae acquire vitamin $B_{\text 12}$ through a symbiotic relationship with bacteria,
\href{https://doi.org/10.1038/nature04056}{Nature {\bf 438},
90--93 (2005)}.

\bibitem{Peaudecerf} F.J. Peaudecerf, F. Bunbury, V. Bhardwaj, M.A. Bees, 
A.G. Smith, R.E. Goldstein, and O. Croze, 
Mutualism Between Microbial Populations in Structured Environments: The Role 
of Geometry in Diffusive Exchanges, 
\href{https://doi.org/10.1103/PhysRevE.97.022411}{Phys. Rev. E {\bf 97}, 022411 (2018)}.

\bibitem{FJPthesis} F.J. Peaudecerf, Ph.D. thesis,
Mathematics, 
University of Cambridge (2017).

\bibitem{PedleyKessler} T.J. Pedley and J.O. Kessler, 
Hydrodynamic phenomena in suspensions of swimming 
microorganisms,
\href{https://doi.org/10.1146/annurev.fl.24.010192.001525}{Annu. Rev. Fluid Mech. {\bf 24}, 313-358 (1992)}.

\bibitem{Dombrowski} C. Dombrowski, L. Cisneros, 
S. Chatkaew, J.O. Kessler, and R.E. Goldstein,
Self-concentration and large-scale coherence in
bacterial dynamics,
\href{https://doi.org/10.1103/PhysRevLett.93.098103}{Phys. Rev. Lett. {\bf 93}, 098103 (2004)}.

\bibitem{ARFM} R.E. Goldstein,
Green algae as model organisms for biological fluid 
dynamics,
\href{https://doi.org/10.1103/PhysRevLett.93.098103}{Annu. Rev. Fluid Dynamics {\bf 47}, 343--375 (2015)}.

\bibitem{mutualrxndiff1} J. Cremer, T. Honda, Y. Tang, 
J. Wong-Ng, M. Vergassola, and T. Hwa,
Chemotaxis as a navigation strategy to boost range
expansion,
\href{https://doi.org/10.1038/s41586-019-1733-y}{Nature {\bf 575}, 658--663 (2019)}.

\bibitem{mutualrxndiff2} H. Malchow, Spatio-temporal
pattern formation in nonlinear non-equilibrium plankton 
dynamics,
\href{https://doi.org/10.1098/rspb.1993.0015}{Proc. R. Soc. B {\bf 251}, 103--109 (1993)}.

\bibitem{mutualrxndiff3} R. Martinez-Garcia, C.E. Tarnita, and J.A. Bonachela, Spatial patterns in 
ecological systems: from microbial colonies to 
landscapes,
\href{https://doi.org/10.1042/ETLS20210282}{Emerg. Top.
Life Sciences {\bf 6}, 245--258 (2022)}.

\bibitem{WuLibchaber} X.-L. Wu and A. Libchaber, 
Particle Diffusion in a Quasi-Two-Dimensional Bacterial Bath, 
\href{https://doi.org/10.1103/PhysRevLett.84.3017}{Phys. Rev. Lett. {\bf 84}, 3017--3020 (2000)}.

\bibitem{Leptos} K.C. Leptos, J.S. Guasto, J.P. Gollub, A.I. Pesci, and 
R.E. Goldstein, 
Dynamics of enhanced trancer diffusion in suspensions of swimming eukaryotic
microorganisms, 
\href{https://doi.org/10.1103/PhysRevLett.103.198103}{Phys. Rev. Lett. {\bf 103}, 198103 (2009)}.

\bibitem{Schnitzer} M.J. Schnitzer,
Theory of continuum random walks and application to chemotaxis, 
\href{https://doi.org/10.1103/PhysRevE.48.2553}{Phys. Rev. E {\bf 48}, 2553--2568 (1993)}.

\bibitem{Lau} A.W.C. Lau and T.C. Lubensky, State-dependent diffusion: Thermodynamic
consistency and its path integral formulation, 
\href{https://doi.org/10.1103/PhysRevE.76.011123}{Phys. Rev. E {\bf 76}, 011123 (2007)}.

\bibitem{Weiss} N.O. Weiss, The Expulsion of Magnetic Flux
by Eddies, 
\href{https://doi.org/}{Proc. R. Soc. A {\bf 293}, 310--328 (1966)}.

\bibitem{Tao} L. Tao, M.R.E. Proctor, and N.O. Weiss, 
Flux expulsion by inhomogeneous turbulence, 
\href{https://doi.org/10.1111/j.1365-8711.1998.t01-1-01957.x}{Mon. Not. R. Astron. Soc. {\bf 300}, 907--914 (1998)}.

\bibitem{Steiner} P.J. Steiner, Ph.D. thesis, Plant Science,
University of Cambridge (2017).

\bibitem{CRC} {\tt https://www.chlamycollection.org/}

\bibitem{pnas} I. Tuval, L. Cisneros, C. Dombrowski, C.W. 
Wolgemuth, J.O. Kessler, and R.E. Goldstein,
Bacterial swimming and oxygen transport near contact 
lines,
\href{https://doi.org/10.1073/pnas.040672410}{Proc. Natl. Acad. Sci. USA {\bf 102}, 2277-2282 (2005)}.

\bibitem{KellerSegel} E.F. Keller and L.A. Segel,
Model for chemotaxis, 
\href{https://doi.org/10.1016/0022-5193(71)90050-6}{J. Theor. Biol. {\bf 30}, 225-234 (1971)}.

\bibitem{Hillesdon} A.J. Hillesdon, T.J. Pedley, and J.O. Kessler,
The development of concentration gradients in a suspension of chemotactic bacteria, 
\href{https://doi.org/10.1016/0022-5193(71)90050-6}{Bull. Math. Biol. {\bf 57}, 299-344 (1995)}.

\bibitem{Douarche} C. Douarche, A. Buguin, H. Salman, and A. Libchaber, 
{\it E. coli} and oxygen: A motility transition,
\href{https://doi.org/10.1103/PhysRevLett.102.198101}{Phys. Rev. 
Lett. {\bf 102}, 198101 (2009)}.

\bibitem{Piskovsky2023} V. Piskovsky, N. Oliveira, 
Bacterial motility can govern the dynamics of antibiotic resistance evolution, 
\href{https://doi.org/10.1038/s41467-023-41196-8}{Nature Communications {\bf 14}, 
(2023)}.

\bibitem{Baig2023} Y. Baig, H.R. Ma, H. Xu, L. You,
Autoencoder neural networks enable low dimensional structure analyses of microbial growth dynamics,
\href{https://doi.org/10.1038/s41467-023-43455-0}{Nat Commun. {\bf 14}(1), 7937 (2023)}.

\bibitem{David2013} L. David, C. Maurice, R. Carmody, D. Gootenberg, J. Button, B. Wolfe, A. Ling, A. Devlin, Y. Varma, M. Fischbach, S. Biddinger, R. Dutton, P. Turnbaugh,
Diet rapidly and reproducibly alters the gut microbiome,
\href{https://doi.org/10.1038/nature12820}{Nature {\bf 505}, (2013)}.

\bibitem{Miles} C.J. Miles, A.A. Evans, M.J. Shelley, and 
S.E. Spagnolie, Active matter invasion of a viscous fluid: 
Unstable sheets and a no-flow theorem,
\href{https://doi.org/10.1103/PhysRevLett.122.098002}{Phys. Rev. 
Lett. {\bf 122}, 098002 (2019)}.

\bibitem{Assante} R. Assante, D. Corbett, D. Marenduzzo, and A. Morozov, 
Active turbulence and spontaneous phase separation in inhomogeneous extensile 
active gels, 
\href{https://doi.org/10.1039/D2SM01188C}{Soft Matter {\bf 19}, 
189--198 (2023)}.


\end{thebibliography}
\end{document}